\documentclass{aa} 
\usepackage{graphicx}
\usepackage{txfonts}

\usepackage{textcomp}
\usepackage{gensymb}
\def\comment#1{\relax}

\newcommand{\SRG}{{\it SRG}\,}
\newcommand{\eROSITA}{{\rm eROSITA}\,}
\newcommand{\SRGeROSITA}{{\it SRG}\slash{\rm eROSITA}\,}
\newcommand{\SwiftBAT}{{\it Swift}\slash{\rm BAT}\,}

\newcommand{\her}{{Her\,X-1}}

\usepackage{color}

\begin{document}

   \title{Observations of Her X-1 in  low states during \textit{SRG}/eROSITA all-sky survey}
    \titlerunning{Her X-1 during \textit{SRG}/eROSITA survey}
    \authorrunning{N.~I.~Shakura et al.}

   \author{ N.~I.~Shakura\inst{1,2}\thanks{nikolai.shakura@gmail.com},
            D.~A.~Kolesnikov\inst{1},
            P.~S.~Medvedev\inst{3,4},
            R.~A.~Sunyaev\inst{3,4},
            M.~R.~Gilfanov\inst{3,4},
            K.~A.~Postnov\inst{1,2}
            \and
            S.~V.~Molkov\inst{3}
        }

   \institute{$^1$ Moscow State University, Sternberg Astronomical Institute, 119234 Moscow, Russia\\
            $^2$ Kazan Federal University, 420008 Kazan, Russia\\
            $^3$ Space Research Institute of the Russian Academy of Sciences, 117997 Moscow, Russia\\
            $^4$ Max-Planck-Institut f\"ur Astrophysik (MPA), Karl-Schwarzschild-Str 1, D-85741 Garching, Germany
            }

    \date{\today}
 
\abstract{ eROSITA (extended ROentgen Survey with an Imaging Telescope Array) instrument onboard the Russian-German `Spectrum-Roentgen-Gamma' (SRG) mission observed the Her X-1/HZ Her binary system in multiple scans over the source during the first and second \SRG all-sky surveys. Both observations occurred during a low state of the X-ray source when the outer parts of the  accretion disk blocked the neutron star from view.
The orbital modulation of the X-ray flux was detected during the low states. We argue that the detected X-ray radiation results from scattering of the emission of the central source by three distinct regions: (a) an optically thin hot corona with temperature $\sim (2-4) \times 10^6\,\mathrm{K}$ above the irradiated hemisphere of the optical star; (b) an optically thin hot halo above the accretion disk; and (c) the optically thick cold atmosphere of the optical star. The latter region effectively scatters photons with energies above 5--6 keV.}

   \maketitle
%

\section{Introduction}
HZ Her/Her X-1 is an intermediate-mass X-ray binary consisting of an accreting neutron star with mass $m_{\rm x} = 1.4\,M_{\odot}$ and spin period $P_{\rm x}\approx 1.24$~s, and a  donor star with mass $m_{\rm o} = 2.0\,M_{\odot}$ \citep{1972ApJ...174L.143T}. The light curve of HZ~Her/Her~X-1 demonstrates clear eclipses because of the high inclination of the binary system to the line of sight, about $90\degree$. The optical star HZ Her fills its Roche lobe \citep{1974ApJ...187..345C} and shows a significant modulation with an orbital period of $P_\mathrm{b}=1.7$ d \citep{1972ApJ...178L...1B}. This modulation is due to a strong irradiation effect of the donor star \citep{1973SvA....17....1L}, as was first found from the inspection of archival photo-plates \citep{1972IBVS..720....1C}. The reflection effect from the heated atmosphere of HZ~Her was also observed in the extreme ultraviolet by the  \textit{EUVE} (Extreme Ultraviolet Explorer) satellite \citep{1999ApJ...521..328L, 2000MNRAS.315..735L, 2003MNRAS.342..446L, 2010ApJ...715..897L} and by the \textit{Astron} satellite \citep{1992SvA....36...41S}. 

Soon after the discovery of the X-ray source in 1971 \citep{1972IAUC.2412....1S, 1972ApJ...174L.143T}, a 35-day  modulation of the X-ray flux was discovered by the \textit{Uhuru} satellite \citep{1973ApJ...184..227G}. For some time, only the Main-on state of the 35-day cycle was known. Using the data obtained by \textit{Copernicus} and \textit{Ariel V} satellites, \citet{fabian_long-term_1973} and \citet{1975Natur.256..712C}, respectively, discovered a lower-amplitude Short-on state. Later on, \citet{1976ApJ...209L.131J} analyzed the archive \textit{Uhuru} observations and revealed the presence of the "Short-on" state in the data. 
Therefore, the 35-day X-ray cycle of Her X-1 comprises four states:
(1) the Main-on lasting approximately seven orbits with the highest X-ray flux;  (2) the first low state lasting approximately four orbits; (3) the Short-on lasting approximately four orbits with the X-ray flux about three times as low as in the Main-on; and (4) the second low-state lasting approximately four orbits (see, e.g., \cite{1998MNRAS.300..992S} for more details and \cite{2020ApJ...902..146L} for a recent update). 
The 35-day cycle is associated with a tilted, retrograde precessing accretion disk. In the middle of the Main-on and Short-on, the disk is 
maximum open to the observer's view, 
and the X-ray source is visible. During the low states, the outer parts of the  tilted disk block the  X-ray source from the observer's view. 
The analysis of a large amount of optical photometric data for HZ Her \citep{1973ApJ...186..617B, 1978pans.proc..111B} supports the presence of such a disk. However, this interpretation is not unique. For example, in the model elaborated by \cite{2002MNRAS.334..847L}, a geometrically thin twisted disk with a thick inner ring and extended central X-ray source is used to explain the ASM/RXTE (All-Sky Monitor of the Rossi X-ray Timing Explorer) light curve. 
This model was also applied to explain orbital modulation of the extreme ultraviolet emission at low states of Her X-1 observed by the \textit{EUVE} satellite \citep{2003MNRAS.342..446L}.
Recent joint observations of Her X-1 by  XMM-\textit{Newton} (X-ray Multi-Mirror Mission)  and NuSTAR (Nuclear Spectroscopic Telescope Array) telescopes confirmed the presence of the warped, retrograde precessing accretion disk \cite{2021arXiv210205097B}.

In 1985, using observations from \textit{EXOSAT} (The European X-ray Observatory SATellite), \citet{1985Natur.313..119P} for the first time detected X-ray flux during both the first and second low states. 
X-ray flux detected by the RXTE/ASM observations of Her X-1 in the low state (Fig. 4, bottom panel, in \cite{1999ApJ...510..974S}) does not show significant orbital modulation outside X-ray eclipses. More detailed RXTE/PCA (RXTE Proportional Counter Array) observations of Her X-1 in the low state starting at MJD 52261.52 measured orbital modulation of the flux \citep{2005MNRAS.361.1393I, 2011ApJ...736...74L, 2015MNRAS.453.4222A}. 
In 2002, \textit{Chandra} detected Her X-1 during its low-state \citep{ji2009}. In 2017, a low-state preceding a turn-on of Her X-1 was observed with the \textit{AstroSat} satellite \citep{2019ApJ...871..152L}.

\citet{1993A&A...273..147M} analyzed a Short-on of Her X-1 using the data obtained by the \textit{ROSAT} (ROentgen SATellite) all-sky X-ray survey. 
Since 2019, a new, deeper all-sky X-ray survey has been carrying out by \(\SRGeROSITA\) mission from the vicinity of the Earth-Moon $L_2$ point
(see \cite{2020arXiv201003477P}
for a description of the \(\SRGeROSITA\) telescope).

\begin{figure}
   \centering
   \includegraphics[width=\columnwidth]{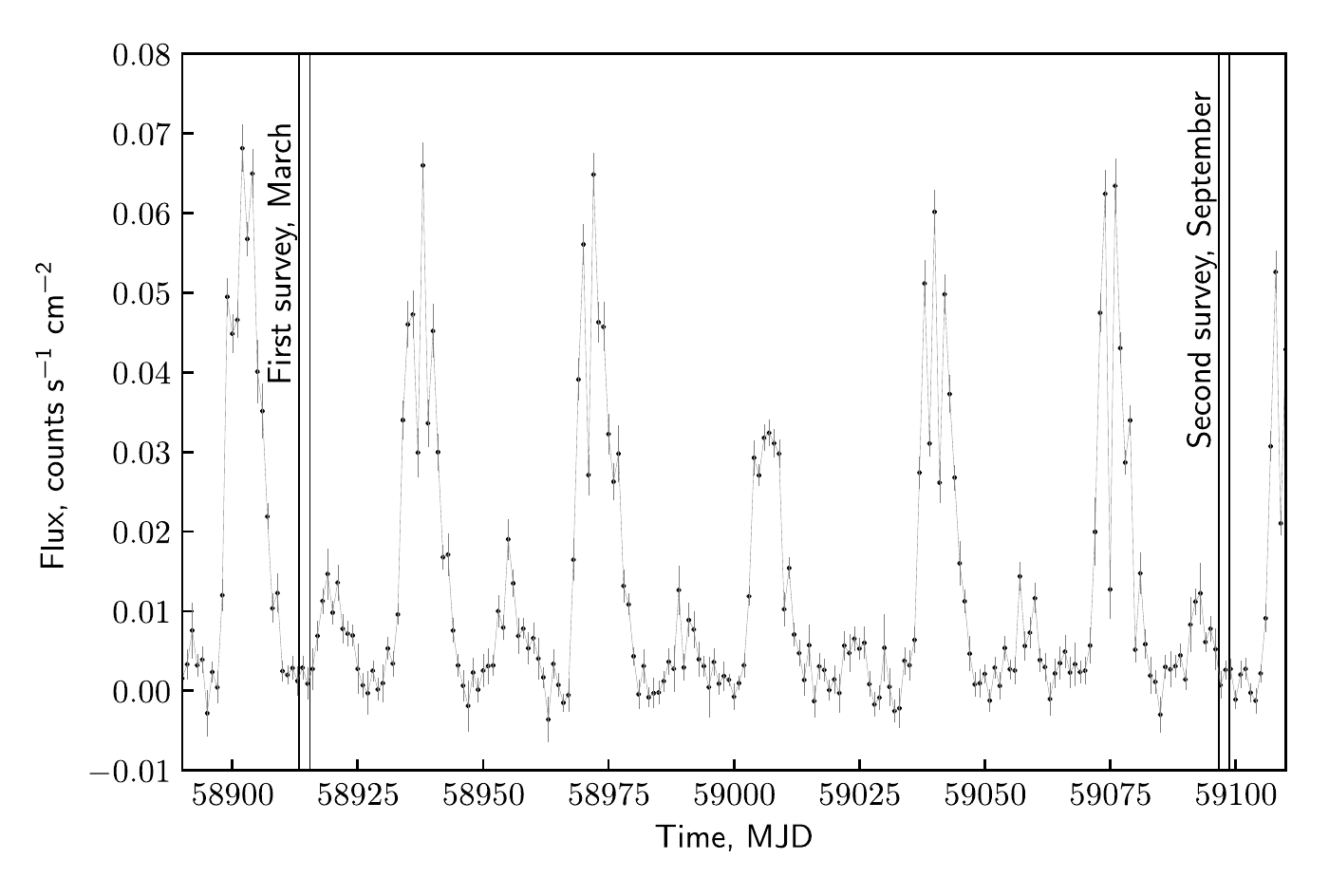}
      \caption{\(\SwiftBAT\)  X-ray flux of Her X-1 as a function of time \citep{2013ApJS..209...14K}. Vertical lines indicate the \(\SRGeROSITA\)  observations of Her X-1. 
              }
         \label{f:fig1}
   \end{figure}

Her X-1 was observed several times at the beginning of March and September 2020. The period of rotation of the \(\SRGeROSITA\) satellite is about four hours. The spin axis of the  satellite precesses around the direction to the ecliptic poles with a rate of about $1\degree \, \rm d^{-1}$.
In the sky scans, Her X-1 was seen in the field of view of the \(\SRGeROSITA\) telescope during 14--16 four-hour intervals, which amounts to 1.5 orbital periods of Her X-1. All the observations of Her X-1 happened during the low state of the X-ray source; see Fig.\ref{f:fig1}.

The \(\SRGeROSITA\) telescope has detected a clear signal from the source modulated with the orbital period of the Her X-1/HZ Her binary. Near the orbital phase zero (the eclipse center), the X-ray flux was minimal. Outside the minimum, the X-ray flux was observed to vary with the orbital phase.

In the present paper, we analyze and model the X-ray flux variation from Her X-1 detected by the \(\SRGeROSITA\) in the low state of the source. The structure of the paper is as follows. In Section \ref{s:Xdata}, \(\SRGeROSITA\) observations  and their spectral modeling are presented. In Section \ref{s:corona}, scattering of X-rays in a hot corona above the irradiated atmosphere of HZ Her is described and calculated. Orbital modulation in the low states of Her X-1 observed by \(\SRGeROSITA\) are computed in Section \ref{s:orb_lc}. In Section \ref{s:cold} we discuss X-ray reflection in the cold photosphere of HZ Her and in Section \ref{s:concl} we formulate our main results. In Appendices A and B, we provide details of the calculation of X-ray scattering in a hot optically thin corona above the illuminated part and photosphere of  the optical star, respectively.

\section{X-ray data}
\label{s:Xdata}

In the first all-sky survey, the position of \her\  was scanned by the \SRG observatory 14 times with a four-hour cadence during 
March 5--7, 2020. 
In these observations, we used data from six \eROSITA cameras, excluding the fifth telescope module (TM).
The net count rate for the source was $3.94\pm0.08$ cts/s in the 0.3--10 keV energy band with a total exposure time of 570 sec. During the second all-sky survey, the source was observed with \(\SRGeROSITA\) on Sep. 4--6, 2020 in a similar manner. All seven TMs were watching the source simultaneously, and the total exposure time was 588 sec. The  net count rate for the source was obtained at the level of $5.8\pm0.1$ cts/s in the 0.3--10 keV energy band.   

The \eROSITA raw data were processed by the calibration pipeline based on the \eROSITA Science Analysis Software System (eSASS) and using pre-flight calibration data. We extracted the source spectra and light curves using a circular region with a radius of 80~arcsec (corresponding to 99\% encircled energy) centered on the source position. An annulus region with inner and outer radii of 100~arcsec and 200~arcsec around \her\  was chosen for the background extraction. We also exclude a circular part with a 15 arcsec radius masking a faint source in the background extraction area.

\comment{
   \begin{figure}
   \centering
   \includegraphics[width=0.9\columnwidth]{figures/xray_img.png}
      \caption{Combined X-ray image of \her\ from two all-sky surveys by \SRG/\eROSITA\ in the 0.3--2 keV energy range smoothed by a Gaussian filter with $\sigma=6^{\prime\prime}$. The position of \her\ was scanned on Mar. 5--7, 2020 and Sep. 4--6, 2020 during the first and the second surveys, respectively. The white solid circle depicts the source extraction region with a radius of 80~arcsec and the yellow dashed circles show the annulus used for background extraction (100 to 200~arcsec). The red circle (15~arcsec) masks a faint source and was excluded from the background extraction region.}
         \label{f:xray_img}
   \end{figure}
}

Although the source is quite bright, it is not bright enough for the photon pile-up to become important. Indeed, the source count rate did not exceed the nominal $\sim 1$ count per second per TM level, which is currently recommended by the calibration team as a threshold signaling that the photon pile-up may be important. The highest full band count rate of $\approx 1.2$ counts/s was achieved in TM5 during the second sky survey, which is compatible with the above threshold. We further checked the source image and spectrum for the pile-up signatures and did not find any. We thus conclude that the source is not subject to significant pile-up effects.

The spectra were extracted for the total exposure in each survey and for each source passage by \(\SRGeROSITA\) (14 spectra per survey). 
We fitted spectra in the 0.3--8 keV energy range.
 For the final spectral analysis (Table~\ref{t:bestfit}, Fig.~\ref{f:surv_spec}--\ref{fig:lightcurves})  we used combined data of all operational telescope modules. The spectra were next rebinned so as to ensure a minimum of five counts per energy bin by means of the standard tool \texttt{GRPPHA}. We then used the C-statistic \citep{cstat} with correction for the background subtraction (``W-statistic'') in the \texttt{XSPEC} package (version 12.11.0, \citealt{Arnaud1996}) to analyze the data. 

\subsection{Average spectra}

We start with analyzing the spectra averaged over all passages in each survey.  The X-ray spectra obtained from the first and second all-sky surveys have similar shapes (see Fig.~\ref{f:surv_spec}), with normalization being higher during the scans in September 2020. The spectra can be characterized by a complex shape at low energies, an almost flat continuum above 2 keV, and a hint of the presence of emission lines. 

Extensive studies of \her\ at different phases of the 35-day cycle by different X-ray satellite have established the broadband continuum emission model involving, in particular, a thermal blackbody component with a temperature of $\approx 0.1$ keV that dominates the spectrum at low energies and a broken power law with an exponential cut-off at higher energies (up to $\sim 200$ keV, see \citealt{mccray1982, 1994ApJ...434..341L, 1995ApJ...450..339L, 1995MNRAS.276..607L, 1997A&A...327..215O, 1997MNRAS.287..622L, 1998A&A...329L..41D, 2001ApJ...547..449L, kuster2005, 2008A&A...482..907K, 2019ApJ...871..152L}). It was also suggested that outside the Main-on state, a partial covering model is needed to account for the spectral shape change \citep{ramsay2002}. 
However, given the limited statistics of our $\approx 200$ sec observations, we use a simplified model to describe the continuum emission by combining a blackbody component and a power law. We also apply the \texttt{TBabs} model \citep{wilms2000} with the neutral hydrogen column density fixed at the Galactic value in the direction to \her, $N_{\rm H}= 1.54 \times10^{20}$~cm$^{-2}$, obtained from the HI4PI map \citep{HI4PI_collab}.

In addition to the complex continuum,  the spectrum of \her\ shows a wealth of X-ray lines. The first spectral features were discovered in observations by BeppoSAX (Satellite per Astronomia X), which revealed a broad Fe L line complex at 0.9--1 keV, the Fe K line complex at 6.4 keV \citep{1997A&A...327..215O, mccray1982, oosterbroek2001}, and also a cyclotron absorption feature at $\sim 40$ keV 
(see \cite{1978ApJ...219L.105T}, \cite{2016A&A...590A..91S}, \cite{2017A&A...606L..13S}, and \cite{2020A&A...642A.196S} for a recent discussion). The low-energy X-ray lines were then studied in depth by
\cite{jimenez2005} and \cite{ji2009} using high-resolution gratings spectra from \textit{Chandra}, which resolved a dozen recombination lines originating from photoionized gas emission. 

\subsection{X-ray spectral modeling}

In this paper, we put forward a model that attributes the photoionized component to the emission from a hot optically thin corona above the donor star (see Section~\ref{s:corona} and Fig. \ref{f:corona}). To investigate the contribution of this component to the \her\ spectrum in the low state, we use \texttt{XSTAR}'s \textsc{photemis} (v. 2.54, \citealt{kallman2001}) semi-analytic model in \texttt{XSPEC}. Using \texttt{XSTAR}, we calculate the ion fractions and atomic level populations file to reproduce relevant physical conditions. In \texttt{XSTAR} run, we assume a spherical geometry of the wind, set the gas density to $10^{12}$ cm$^{-3}$, and fix elemental abundances  to the solar values which are defined in \texttt{XSTAR} relative to \cite{grevesse1996}. For the illuminating source, we set the luminosity $2 \times 10^{37}$ erg/s and a power-law spectrum with photon index $\Gamma=1$, which roughly corresponds to the central source in \her. We set the initial gas temperature at $T_{init} = 10^5$ K and allow \texttt{XSTAR} to calculate it assuming the thermal equilibrium of gas. Due to insufficient counts, we do not try to fit the ionization parameter value and fix it at $\xi = L_i / n R^2 = 1000$ (see Section~\ref{s:corona}), where $L_i$ is the ionizing luminosity, $R$ is distance from the ionizing source, and $n$ is the gas number density. This photoionization parameter value follows from theoretical calculations of induced stellar wind from X-ray illuminated atmosphere of HZ Her \citep{1973Ap&SS..23..117B, 1974A&A....31..249B}. Then the equilibrium gas temperature  is $T_{eq} \approx 2\times 10^6$ K.

\begin{figure*}
  \centering
  \includegraphics[width=0.8\textwidth]{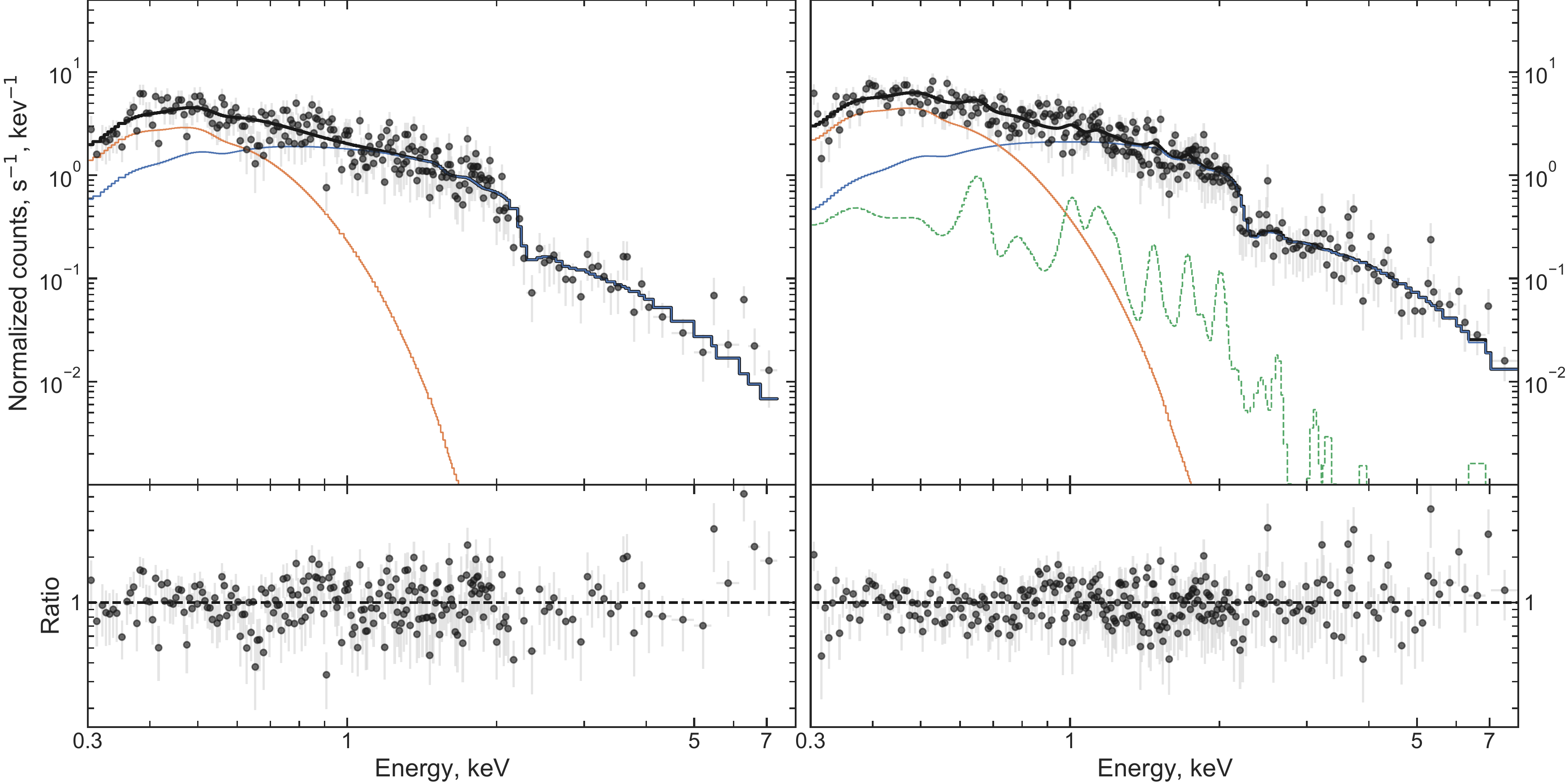}
      \caption{\eROSITA spectra of Her X-1 obtained during the first and the second sky surveys (left and right panels, respectively). The solid black line shows the best-fit model (see Table~\ref{t:bestfit}), and the green dashed line shows the contribution from the proper emission of hot photoionized plasma (not detected in the first survey). Thin red and blue lines show the blackbody and power-law components, respectively. The bottom panels show the ratio of the data to the folded model. The visible excess in the residuals at higher energies is due to the Fe\,K complex and possible reflection 
      from the cold photosphere of the optical star, which are  unaccounted for in the spectral modeling (see Section \ref{s:cold}).}
         \label{f:surv_spec}
  \end{figure*}

We thus fit the broadband (0.3--8 keV) spectra using the combined model which reads in \texttt{XSPEC} as  \textsc{TBabs (bbodyrad + powerlaw + photemis)}.  Figure~\ref{f:surv_spec} shows the best-fit model and the contribution of various components to the \eROSITA\ spectra. The best-fit parameters and the 90\% confidence ranges are listed in Table~\ref{t:bestfit}. We found no evidence for the photoionization component in the spectrum from the first survey with a 90\% upper limit on its emission measure at $8\times10^{56}$ cm$^{-3}$. For the second survey, we found a marginal improvement of the fit ($\delta$C-statistic= \textminus 6 for $\delta$d.o.f.= $+1$), which corresponds to a statistical significance of $\approx 2.6 \sigma$. We found no significant variation in the blackbody temperature and normalization between the averaged spectra from the first and the second surveys. The found best-fit values are in good agreement with the \textit{Chandra} results for observations in the low state of Her X-1 \citep{ji2009}. 

\begin{table}
\caption{Best-fit parameters obtained from fitting to the survey-averaged energy spectrum of \her\ from the first and the second \SRG\ all-sky surveys. Errors are quoted at the 90\% confidence level.
}
\centering
\begin{tabular}{lcc}
\hline\hline
Parameters &  First survey & Second survey \\
\hline
$N_H$, cm$^{-2}$ & $1.54\times10^{20}$ $^*$ &  $1.54\times10^{20}$ $^*$ \\
 $kT_{\mathrm{bb}}$, eV & $103_{-12}^{+11}$ &  $106 \pm 9$ \\
 $R_{\mathrm{bb}}$\tablefootmark{a}, km & $41.3_{-7.9}^{+11.8}$ & $46.0_{-7.5}^{+9.6}$ \\
$ \Gamma_{\mathrm{pow}} $ & $1.13_{-0.17}^{+0.16}$ & $0.71_{-0.14}^{+0.13}$ \\
$K_{\mathrm{pow}}$\tablefootmark{b}, & 
$2.22\pm0.25\times10^{-3}$ &  $2.38_{-0.29}^{+0.30}\times10^{-3}$  \\
ph keV$^{-1}$ cm$^{-2}$ s$^{-1}$\\
$ EM_{\mathrm{photemis}}$\tablefootmark{c}, cm$^{-3}$ & $<8 \times 10^{56}$ & $
2.79_{-1.77}^{+1.85} \times 10^{57}$  \\
$\xi$ & $1000^{*}$ & $1000^{*}$ \\
 C-stat\,/\,d.o.f. & {277  / 237} &  {290 / 298}\\
\hline
\end{tabular}
\tablefoot{ 
We use typical interstellar abundances by \cite{wilms2000} for the \textsc{tbabs} component, while for the \textsc{photemis} component elemental abundances  
are fixed to the solar values relative to \cite{grevesse1996}. The parameters $R_{\mathrm{bb}}$ and $EM_{\mathrm{photemis}}$ are calculated assuming the distance of 6.6 kpc to \her \citep{reynolds97}. \\ 
\tablefoottext{a}{
The size of the thermal region derived from the thermal blackbody component normalization: $R_{km}=\sqrt{\mathrm{norm}_{bb}} \times D_{10}$.} 
\tablefoottext{b}{Normalization of the power-law component at 1 keV.}
\tablefoottext{c}{The emission measure of the gas obtained from the \textsc{photemis} best-fit normalization: $EM = 10^{10} 4 \pi D^2 \times \mathrm{norm}_{photemis}$} \tablefoottext{*}{Parameter is fixed during the fit.}
}
\label{t:bestfit}
\end{table}

\subsection{X-ray orbital light curve}

Due to a minor contribution from the photoionized emission (see Fig. \ref{f:surv_spec}), to construct the X-ray orbital light curve during the first and the second survey, we simplify the spectral model by eliminating the photoionization component. We also fix the blackbody component temperature at 0.1 keV (see Table \ref{t:bestfit}). Thus, the broadband spectral model in the low state of Her X-1 can be fitted by three parameters: the power-law and blackbody component normalizations and the power-law photon index. The obtained model makes it possible to investigate the evolution of the continuum components during individual scans, for which the total number of counts is low, namely about $100$. We note that the X-ray spectrum of Her X-1 was described by the blackbody and power-law components for the first time by \cite{mccray1982} using the \textit{Einstein} observations. The blackbody radiation is emitted by the inner parts of the accretion disk. The power-law tail is produced near the surface of the accreting magnetized neutron star. During the low state, the accretion disk and the inner parts of the  neutron star are obscured from the view of the observer by the  outer parts of the warped disk. Both spectral components are observed directly at the Main-on and Short-on states of the 35-day cycle. 

The obtained X-ray flux of the power-law component, dominating in the energy range 0.2--8 keV, as a function of time during the first and the second \SRG\ all-sky surveys is shown in Fig.~\ref{fig:lightcurves} by dots with error bars. The power-law component dominates over the blackbody one above $\sim 0.7$~keV, and therefore below, we use the power-law component as a proxy for the observed X-ray flux in the low state of Her X-1.

\section{Scattering in a hot optically thin corona above the donor star}
\label{s:corona}

\citet{1973Ap&SS..23..117B} and\cite{1974A&A....31..249B} 
were the first to show that, in an X-ray binary with an optical donor star, above the irradiated side of the donor, a hot ($T\sim (2-4) \times 10^{6}\, \rm K$) corona is formed that is optically thin to scattering free electrons ($\tau_{\rm s} \lesssim 0.05 - 0.1$).
The gas number density of the corona is $n\sim 10^{12}\, \rm cm^{-3}$. The absorption optical depth in the corona is negligible ($\tau_{\rm a} \ll 1$).  

Following \citet{1981ApJ...249..422K}, the photoionization parameter $\Xi$ is defined as  \:
\begin{equation}
    \Xi = \frac{F}{n c k_{\rm B} T} = \frac{L_{\rm x}}{4 \pi r^2 n c k_{\rm B} T} = \frac{\xi}{4 \pi c k_{\rm B} T} \approx 20000 \frac{\xi}{T}.
\end{equation}
The value of $\Xi$ in the corona is about 5--10. Along with thermally stable states, there are thermally nonstable states of the corona \citep[see, e.g.,][]{2016A&A...596A..65M}. Therefore, we expect thermal stratification of the coronal plasma and hence time variations of the scattered X-ray flux, which is indeed observed (see Fig. \ref{fig:lightcurves}). Further away from the stellar surface, the corona is transformed into a hot stellar wind with a temperature of about $10^7\, \rm K$. 

In Appendix \ref{app1}, the scattered X-ray flux from such a corona illuminated by a point-like central X-ray source is computed analytically using the single scattering approximation. The calculations presented in Appendix A suggest that the scattered X-ray luminosity of the corona is about $2 \times 10^{35}\,\rm{erg\,s^{-1}}$. The proper  X-ray luminosity of the hot photoionized coronal plasma is much smaller, of the order of $3 \times 10^{34}\, \rm{erg\,s^{-1}}$. The characteristic time of emission is approximately some tens of seconds.


In the case of Her X-1, accurate computation of scattering from the hot corona above the X-ray-illuminated atmosphere of HZ Her requires taking into account the shadow from the tilted, twisted accretion disk around the neutron star and the complicated geometry of the corona. In our calculations,   
we use several simplifying approximations.
\begin{enumerate}
    \item 
    The corona is assumed to have a constant density and temperature; its size is equal to that of exponential atmosphere,  $H = k_{\rm B} T R^2 / m_{\rm p} G M$. Depending on the temperature, $H \sim 0.1$--$0.3 R$. The corona is virtually absent inside the X-ray shadow produced by a warped accretion disk around the central source; see Fig. \ref{fig:lightcurves}.

    \item
    The observed X-ray flux is assumed to be proportional to the visible volume of the hot irradiated plasma. The corona can be partially screened from the observer by the star itself and by the accretion disk.

    \item 
    Each volume element $dV$ of the corona is assumed to scatter photons independently from others. The flux scattered by a volume element is proportional only to the incident X-ray flux $F = L_{\mathrm{x}} / 4 \mathrm{\pi} r^2$, where $r$ is the distance between the volume element $dV$ and the central X-ray source with luminosity $L_{\mathrm{x}}$.

\end{enumerate}

In a homogeneous corona, the specific luminosity of single-scattered X-ray radiation from the corona reads:
\begin{equation}
\label{eq:int1}
    \frac{dL}{d\Omega} = \int n \sigma_{\rm T} \frac{L_{\mathrm{x}}}{4 \mathrm{\pi} r^2} dV,
\end{equation}
where the integral is taken over the volume visible from a given direction, 
$x(\gamma)$ is the scattering diagram, $\gamma$ is the angle between the given direction and the line connecting the volume element $dV$ and X-ray source, 
$\sigma_{\rm T}$ is the Thomson cross-section, and $n$ is the number density of electrons in the volume element $dV$.

The integral in Eq. (\ref{eq:int1}) is calculated numerically using the following procedure. The corona is approximated by $N$ fixed randomly distributed  points filling the layer with height $H$ above the donor star's irradiated surface; see Fig. \ref{fig:lightcurves}. Each point is checked for whether it is blocked from the view of the observer by the donor star or the accretion disk or falls within the disk's X-ray shadow region. For points which are not blocked and are not inside the shadow, the distance $r$ to the X-ray source is calculated. Then, the integral (\ref{eq:int1}) can be approximated as: 
\begin{equation}
    \frac{dL}{d\Omega} = \frac{V}{N}\sum_{i=1}^{N} \delta_i n \sigma_{\rm T} \frac{L_{\rm x}}{4 \pi r_i^2},
\end{equation}
where $V$ is the full volume of the corona including the X-ray shadow and blocked regions. The volume $V$ is computed numerically. $\delta_i = 0$ if point $i$ is blocked from the view of the observer by the star or the accretion disk, or is inside the disk's X-ray shadow region, and $\delta_i = 1$ if the point $i$ is not blocked, nor falls inside the shadow region.   

\begin{figure}
         \centering
         \includegraphics[width=\columnwidth]{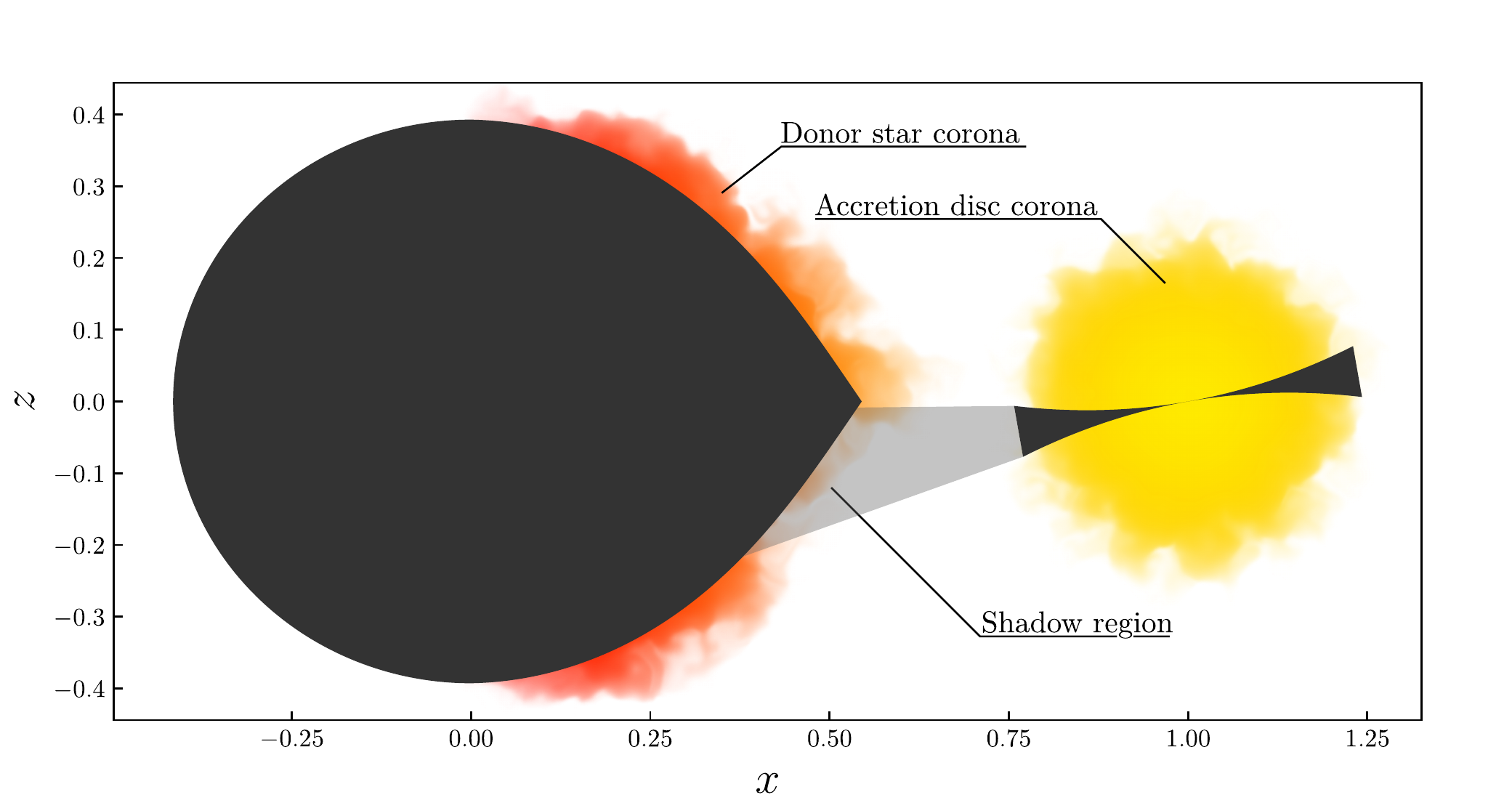}
         \caption{Schematic geometrical model of Her X-1/HZ Her used to calculate the X-ray orbital light curve. The model includes the Roche-lobe filling star (to the left), a hot corona above its X-ray-illuminated part (in orange), the accretion disk around the neutron star (to the right), and a hot halo above the accretion disk (disk corona, in yellow). Also shown is the accretion disk shadow on the illuminated part of HZ Her.}
        \label{f:corona}
     \end{figure}

  \begin{figure}
  \centering
  A\\
  \includegraphics[width=\columnwidth]{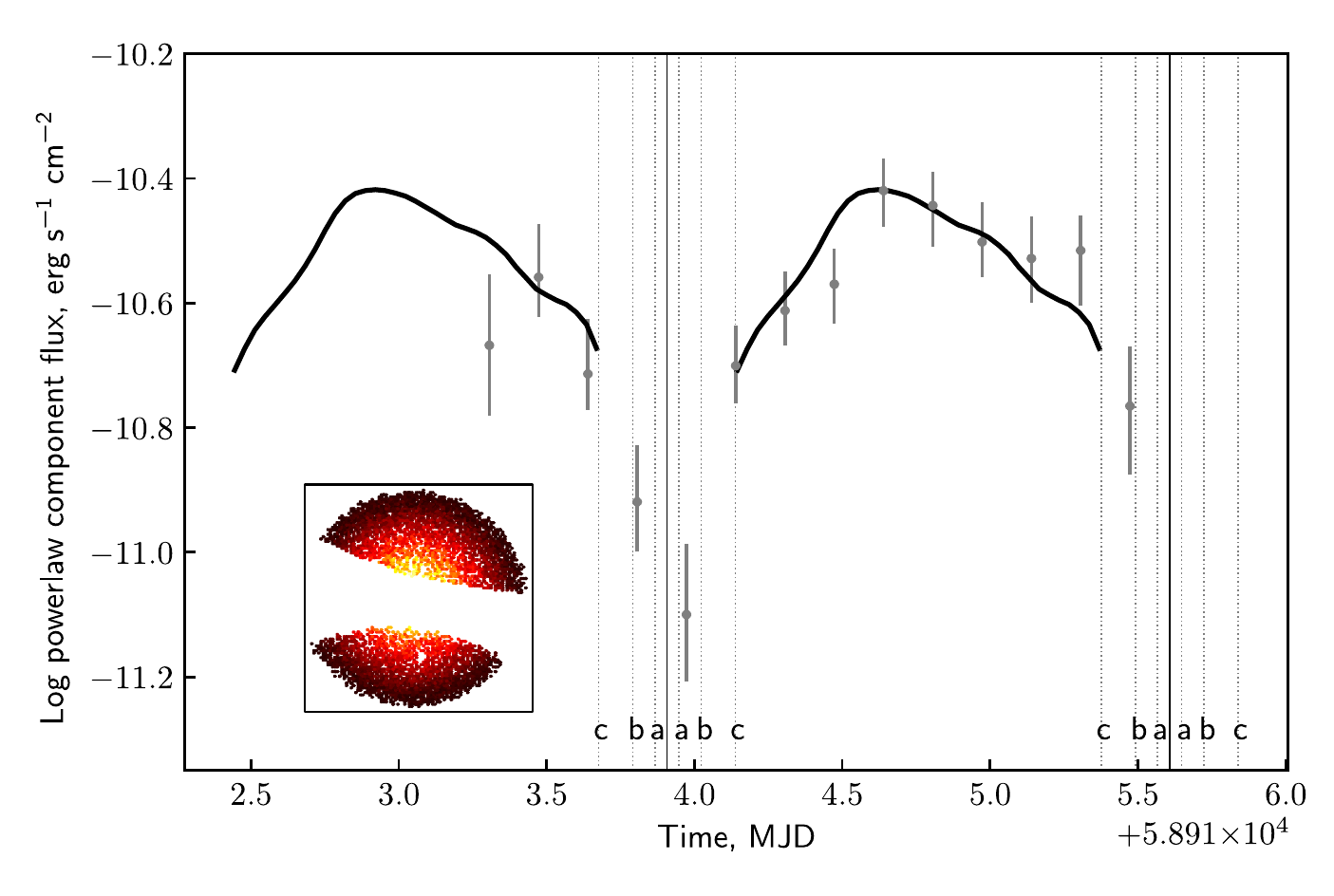}
  
  B\\
  \includegraphics[width=\columnwidth]{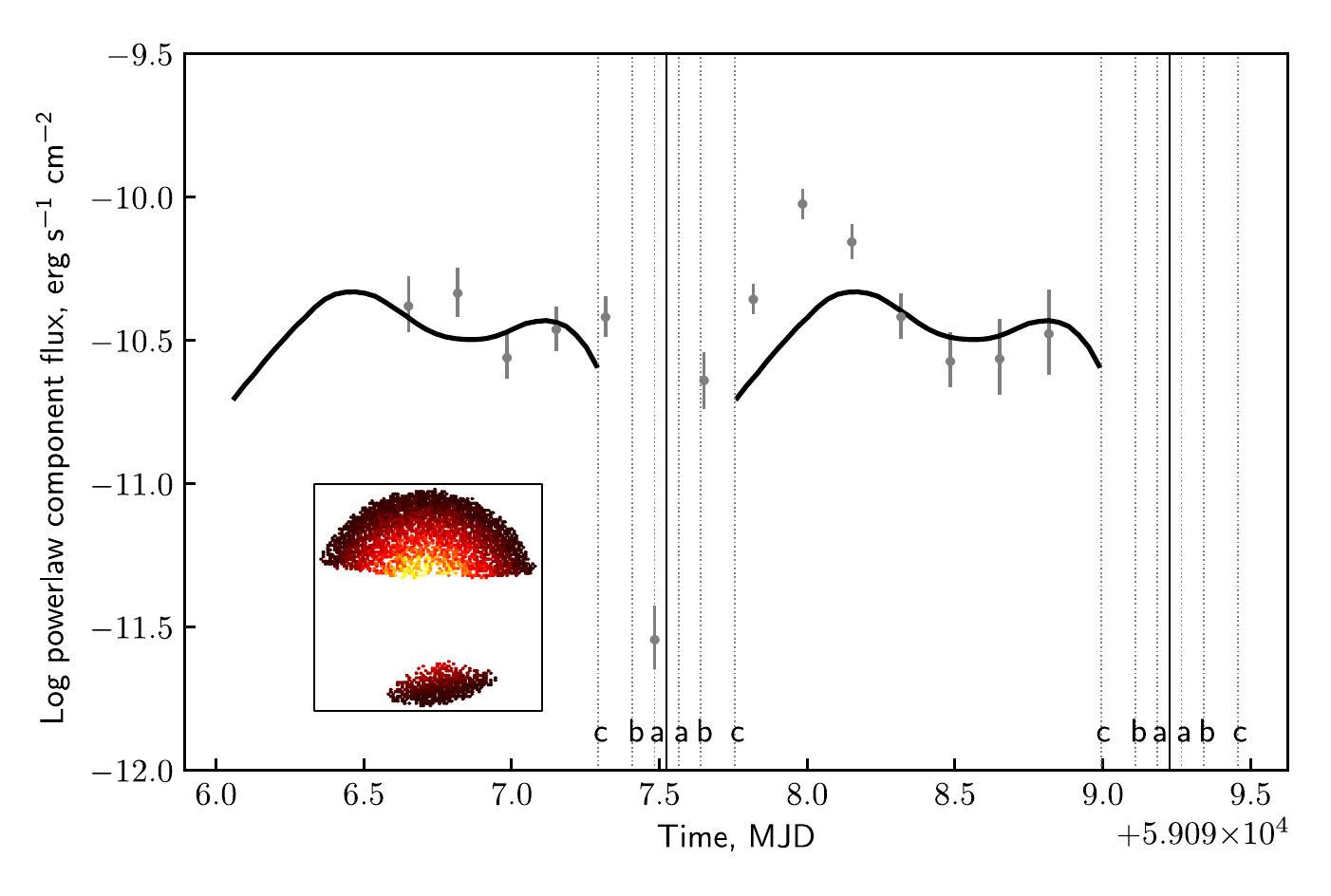}
      \caption{X-ray flux of the power-law component in the energy range 0.2 -- 8 keV from Her X-1 as a function of time during the first (A) and second (B) survey. The solid line is the theoretical flux calculated as described in Section \ref{s:orb_lc}. Figures inside plots show the brightness of scattered X-rays in the corona at the orbital phase 0.5. The white region is the X-ray shadow.  The solid vertical lines indicate the orbital phase zero. The dotted vertical lines a, b, and c before and after the orbital phase zero indicate different accretion disk ingress and egress moments. Here, c--a and a--c intervals before and after phase zero, respectively, indicate the orbital phases where the donor star partially covers the accretion disk. Line b indicates expected moments of ingress and egress of the X-ray source. The a--a interval covers the total eclipse of the disk by the donor star.}
    \label{fig:lightcurves}
\end{figure}

\begin{table*}
\caption{Model parameters during first and second survey}
    \centering
    \begin{tabular}{lcc}
    \hline
    \hline
    Parameters & First survey & Second survey \\
    \hline
    Mass ratio $q = m_{\rm x}/m_{\rm o}$ & $0.6448$\tablefootmark{a} & $0.6448$\\
    Inclination $i$ & $87\degree$ & $87\degree$\\
    Relative outer edge semi-width $h_{\rm out}/R_{\rm o}$ & $0.15$ & $0.15$ \\ 
    Disk outer edge tilt $\theta_{\rm out}$ & $15\degree$ & $18\degree$ \\
    Disk inner edge tilt $\theta_{\rm in}$  & $10\degree$ & $3\degree$ \\
    Disk phase angle $\Phi$ & $-90\degree$ & $-210\degree$ \\
    Disk twist $Z = \varphi_{\rm out} - \varphi_{\rm in}$ & $40\degree$ & $-90\degree$ \\
    Disk halo constant relative additive to flux & $0.3$ & $0.1$ \\
    Corona height $H/R_{\rm opt}$ & $0.15$ & $0.15$ \\
    \hline
    \end{tabular}
    \label{tab:binary_parameters}
\tablefoot{
\tablefoottext{a}{\cite{2014ApJ...793...79L}}
}
\end{table*}

\section{Orbital X-ray light curve in the low state of Her X-1 }
\label{s:orb_lc}

As shown in the previous section, X-ray emission in the low-state of Her X-1 is dominated by the scattered radiation in the hot corona above the illuminated atmosphere of HZ Her. 
To calculate the orbital X-ray light curve, we use the geometrical model of a binary system with a Roche-lobe-filling star and a tilted, warped accretion disk around the compact star, as implemented in the numerical code developed in \citet{2020MNRAS.499.1747K} and supplemented here with a scattering X-ray corona above the illuminated atmosphere of  the optical star
(see Fig. \ref{f:corona}).

A hot halo above the accretion disk itself (the disk corona) also scatters X-rays from the central source (see, e.g., \cite{1989SvA....33..638B,1992IAUS..151..449B}). \textit{Ginga}/LAC observations of Her X-1 during a short-on state suggested the presence of an extended X-ray source \citep{2000MNRAS.315..735L, 2019ApJ...871..152L}. Later on, evidence for a hot disk corona in Her X-1 was found from RXTE observations  \citep{2015ApJ...800...32L}. In our modeling, we assume that the hot accretion disk corona constantly adds to the total flux at all orbital phases except for eclipses of the central X-ray source and the accretion disk (0.0--0.13 and 0.87--1.0). To calculate the X-ray light curve at the eclipse phases, one should specify the accretion disk corona structure. This can be quite complicated (see, e.g.,  the modeling of orbital X-ray light curve with a nonconstant density disk corona in \cite{2015ApJ...800...32L}). In an almost edge-on binary like Her X-1, some orbital modulation from the structured accretion disk corona can be due to eclipses of the corona's inner parts by the outer boundary of the twisted tilted accretion disk. However, the quality of X-ray data under analysis does not allow us to test such additional effects to scattering on the coronal plasma above the illuminated atmosphere of HZ Her. 
A continuous set of dedicated observations during the X-ray eclipse is needed to probe the central X-ray source's hot corona. This is a different problem, and in this paper we exclude the eclipse phases (0.0--0.13 and 0.87--1.0) from our analysis.

The model X-ray light curve of Her X-1 in the low states observed during the first and second \eROSITA surveys is presented in Fig. \ref{fig:lightcurves} (A and B, respectively). The solid curves indicate the calculated X-ray flux scattered in the hot corona above HZ Her and accretion disk while taking into account shadows from the precessing, tilted, warped accretion disk. 
On the left panels of the plots, the X-ray brightness of the corona above the illuminated atmosphere of HZ Her (in color) and the disk shadows (in white) are shown for the orbital phase 0.5.
The orbital ephemeris of Her X-1 is taken from \cite{2009A&A...500..883S}. The disk corona's contribution is presented in Table \ref{tab:binary_parameters} in units of scattered flux from the donor star's corona. Dots with vertical bars show the observed \eROSITA flux. We assume that the X-ray luminosity of the central source remained constant during both \eROSITA observations.

Observations A and B occurred at different phases of the 35-day cycle of Her X-1. The first (A) and second (B) survey observations happened in the low states preceding and following the Short-on. According to the \(\SwiftBAT\) (Swift Burst Alert Telescope) X-ray data (see Fig. \ref{f:fig1}), the 35-day phase of the middle of \eROSITA observations is $\Phi \approx 0.5$ during the first survey and $\Phi \approx 0.8$ during the second survey. The difference is clearly seen in the observed and calculated light curves that are mostly shaped by the form of the precessing disk shadow shielding the corona from the central X-ray source.

Figure \ref{fig:lightcurves} suggests that the modulation of the X-ray emission observed during both the first and second observations of Her X-1 can be explained by scattering in the hot corona above the illuminated atmosphere of HZ Her with an account of shadows produced by the precessing, tilted, warped accretion disk around the central neutron star  \citep{2020MNRAS.499.1747K}. 

\section{X-ray absorption and scattering in the cold optically thick donor's photosphere}
\label{s:cold}

Above, we calculate the X-ray light curve from Her X-1 in the low state 
in the energy range where the scattered radiation from the hot corona above HZ Her dominates. At high photon energies, the X-ray reflection from the cold photosphere of HZ Her should also be taken into account \citep{1974A&A....31..249B}. Evidence for hard X-ray reflection from HZ Her during the low state of Her X-1 was found from RXTE/PCA observations by \cite{2015MNRAS.453.4222A}.

Indeed, as shown in Appendix \ref{app2}, in the hot corona above the optical star, the photon scattering optical depth $\tau_{\rm s} \lesssim 0.1$, which means that only $\lesssim 10$\% of incident photons can be scattered.
The rest of the photons are absorbed and partially reflected by the photosphere of the cold donor. 
The absorption cross-section in the cold plasma strongly depends on the photon energy, $\sigma_{\rm a} \sim E^{-3}$. The single scattering albedo is approximately $\lambda \approx 1/2$ at the energy 6--8 keV. Above 8 keV, about half of all photons are scattered by the cold plasma and escape from the photosphere. The exact intensity $I(\tau = 0, \mu, \phi)$ of scattered photons in the cold photosphere is calculated in Appendix \ref{app2}.

As seen from Fig. \ref{fig:qq} in Appendix \ref{app2}, above 8 keV, the X-ray reflection from the cold photosphere of HZ Her could add $\sim 30\%$ to the total flux. The increase in the spectral residuals at high energies visible in the lower panels of Fig. \ref{f:surv_spec} could be attributed to this reflection.

\section{Discussion and Conclusion}
\label{s:concl}

In this paper, we present the \(\SRGeROSITA\) X-ray observations of Her X-1 during two low-states of the 35-day cycle. Observations were performed during the first and second surveys in March and September 2020, and each covered about 1.5 binary orbital periods. Orbital modulation of X-ray flux during the low-states was clearly detected (see Fig. \ref{fig:lightcurves}).

In both observations, the X-ray spectrum of the source can be fitted by a blackbody and power-law component, with only a minor contribution of proper emission from the hot photoionized plasma (see Fig. \ref{f:surv_spec} and Table \ref{t:bestfit}). The power-law component dominates above $\sim 0.7$~keV. The observed short-time variability could be due to thermal instability in the photoionized plasma (see, e.g., \cite{2016A&A...596A..65M}). Indeed, variable ionized mass outflows  from Her X-1  are observed originating from the nonstationary hot corona above the accretion disk and optical star \citep{2020MNRAS.491.3730K}.  

In the energy range 0.2--8 keV, the orbital modulation from Her X-1 in the low state can be explained  (the solid curve in Fig. \ref{fig:lightcurves}) by a geometrical model of the binary with precessing tilted accretion disk around the central X-ray source \citep{2020MNRAS.499.1747K} supplemented with the X-ray scattering from a hot optically thin corona above the illuminated photosphere of the optical star. At energies close to 8 keV, photons reflection from the cold photosphere of the optical star  should be taken into account (see Appendix B). This could explain the excess of high-energy photons visible in the spectral residuals in the lower panel of Fig. \ref{f:surv_spec}.

The X-ray orbital modulation was detected at the end of the low state during one binary period preceding the Short-on in RXTE/PCA observations \citep{2005MNRAS.361.1393I,2015MNRAS.453.4222A}. In \cite{2015MNRAS.453.4222A}, the X-ray spectra are modeled by three components: (1) a power-law component with photon index $\Gamma \approx 1$ reflected from the donor star photosphere and strong absorption $N_{H} \approx 10^{23}-10^{24}\, \mathrm{cm}^{-2}$, (2) an unabsorbed power-law component with the same $\Gamma$ due to scattering in a hot corona above the accretion disk, and (3) the Fe $K_{\alpha}$ line. In the model of these latter authors, the orbital modulation is due to variable absorption of the hard reflected component from the cold photosphere, while the scattered component from the disk corona remains constant. 

The picture is different in the case of \eROSITA observations. The \eROSITA 0.3--7 keV spectra (see Table \ref{t:bestfit} and Fig. \ref{f:surv_spec}) are best fitted by one power-law component with $\Gamma \approx 1$ and $N_{H} \approx 10^{20}\,\mathrm{cm}^{-2}$ due to scattering in a hot plasma above the irradiated photosphere of HZ~Her, a black-body component with $kT \approx 0.1\,\mathrm{keV}$ due to scattering of the emission from the X-ray-irradiated inner disk \citep{1982ApJ...262..301M}, and a minor contribution from photoionized optically thin plasma. There are traces of the Fe K complex. We stress that in our model, we do not need strong absorption of the reflected X-ray emission from HZ~Her photosphere by cold material along the line of sight. The spectra obtained during short \eROSITA scans with limited count statistics also do not require a partial-covering cold absorber with $N_{H} \sim 10^{22}\, \mathrm{cm}^{-2}$ \citep{2019ApJ...871..152L}. 
We suggest that the observed orbital modulation of the X-ray emission in the low state is due to the changing view of the scattering hot plasma above the irradiated photosphere of HZ~Her partially shadowed by the tilted, twisted accretion disk.

Our analysis excluded the orbital phases around the X-ray eclipse (0.0--0.13 and 0.87--1.0). However, these phases are very interesting because they can be used to probe the structure of the hot corona above the accretion disk (see the analysis of RXTE/PCA eclipses in \cite{2015ApJ...800...32L}).

We conclude that one of the best-studied X-ray binaries, Her X-1/HZ Her, continues to reveal intriguing physical properties that can help us to understand the complicated structure of accretion flows around magnetized neutron stars. The first \eROSITA observations have already significantly contributed to our improved understanding of this problem.

\begin{acknowledgements}
The authors thank the anonymous referee for useful notes that helped us improve the presentation of the results.
The research is supported by the RFBR grant no. 18-502-12025 and the Interdisciplinary Scientific and Educational School of Moscow University' Fundamental and Applied Space Research'. PM acknowledges the hospitality of the Max-Planck Institute for Astrophysics, where part of this work was done. This work is based on observations with \eROSITA telescope onboard \textit{SRG} observatory. The \textit{SRG} observatory was built by Roskosmos in the interests of the Russian Academy of Sciences represented by its Space Research Institute (IKI) in the framework of the Russian Federal Space Program, with the participation of the Deutsches Zentrum für Luft- und Raumfahrt (DLR). The \(\SRGeROSITA\) X-ray telescope was built by a consortium of German Institutes led by MPE, and supported by DLR.  The \textit{SRG} spacecraft was designed, built, launched, and is operated by the Lavochkin Association and its subcontractors. The science data are downlinked via the Deep Space Network Antennae in Bear Lakes, Ussurijsk, and Baykonur, funded by Roskosmos. The \eROSITA data used in this work were processed using the eSASS software system developed by the German \eROSITA Consortium and proprietary data reduction and analysis software developed by the Russian \eROSITA Consortium.
\end{acknowledgements}

%
%


\begin{appendix}

\section{X-ray scattering in an optically thin corona above the donor star}
\label{app1}

\begin{figure}
        \includegraphics[width=\columnwidth]{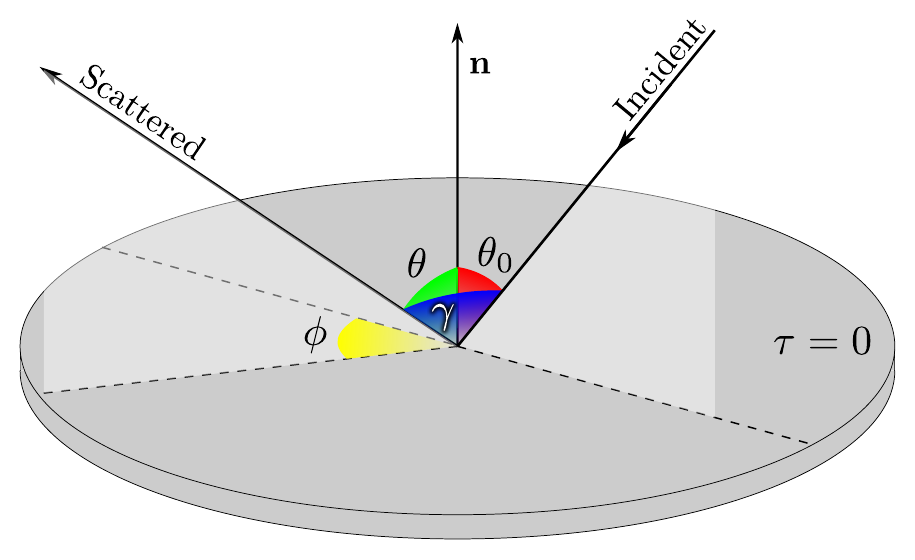}
    \caption{Scheme of scattering of radiation by the semi-infinite plane-parallel atmosphere. $\mathbf{n}$ is the normal vector to the surface, $\theta_0$ is the angle between the direction of the incident ray and normal vector $\mathbf{n}$, $\theta$ is the angle between scattered ray and normal vector $\mathbf{n}$, $\phi$ is the azimuth angle of scattered ray,  and $\gamma$ is the angle between the incident ray and scattered ray.}
    \label{fig:atmosphere}
\end{figure}

The radiation intensity $I(\tau, \mu, \phi)$ in a plane-parallel atmosphere with absorption and scattering obeys the radiative transfer equation, see Fig. \ref{fig:atmosphere}:
\begin{equation}
    \mu \frac{dI(\tau,\mu,\phi)}{d\tau} = I(\tau,\mu,\phi) - S(\tau,\mu,\phi),
        \label{eq:transfer}
\end{equation}
where $\mu = \cos{\theta}$, $\tau$ is the optical depth, $d\tau = -(\kappa + \sigma)dz$, $\kappa$ is the absorption coefficient, and $\sigma$ is scattering coefficient.

The source function $S(\tau,\mu,\phi)$ is defined as follows
\begin{equation}
    S(\tau, \mu, \phi) = S_{\mathrm{inc}} (\tau, \mu, \phi) + \frac{\lambda}{4\pi} \int I(\tau, \mu', \phi') x(\mu, \phi, \mu', \phi') d\mu' d\phi',
   \label{eq:S}
\end{equation}
where the integration is taken over all directions, $\lambda = \sigma/(\kappa + \sigma)$ is the single-scattering albedo, and $x(\mu, \phi, \mu', \phi')$ is the scattering diagram. In the case of random scatterings, the scattering diagram depends only on the angle between the incident and scattering direction, $x(\cos{\gamma})$, 
\begin{equation}
    \cos{\gamma} = \mu\mu' + \sqrt{1-\mu^2}\sqrt{1-\mu'^2}\cos{(\phi-\phi')}
.\end{equation}
The first part in Eq. \ref{eq:S} allows for single-scattering of the incident radiation:
\begin{equation}
    S_{\mathrm{inc}}(\tau, \mu, \phi) = \frac{\lambda}{4 \pi} \int I_{\rm inc}(\tau, \mu', \phi') x(\mu, \phi, \mu', \phi') d\mu' d\phi',
    \label{eq:S_inc}
\end{equation}
where integration is taken over all directions, $I_{\rm inc}$ is the intensity of the direct radiation:
\begin{equation}
    I_{\rm inc}(\tau, \mu, \phi)= F e^{-\tau/\mu_0} \delta(\mu - \mu_0) \delta(\phi),
    \label{eq:I_inc}
\end{equation}
$F$ is the illumination flux perpendicular to the unit surface area, $\mu_0 = \cos{\theta_0}$, $\theta_0$ is the angle between the normal to the surface and incident ray, and $\delta(...)$ is the Dirac delta function.   
By substituting \ref{eq:I_inc} into  \ref{eq:S_inc}, the integral \ref{eq:S_inc} can be calculated to give:
\begin{equation}
    S_{\rm inc}(\tau, \mu, \phi) = \frac{\lambda F}{4\pi} x(\cos{\gamma_0}) e^{-\tau/\mu_0},
    \label{eq:S_inc_2}
\end{equation}
where
\begin{equation}
    \cos{\gamma_0} = \mu\mu_0 + \sqrt{1-\mu^2}\sqrt{1-\mu_0^2}\cos{\phi.}
\end{equation}

With known source function $S(\tau, \mu, \phi)$, the intensity of the escaped radiation reads \citep{1975lpsa.book.....S, 2001nagirner}
\begin{equation}
    I(\tau=0, \mu, \phi) = I_0 = \int_{0}^{\tau_0} S(\tau, \mu, \phi) e^{-\tau/\mu} \frac{d \tau}{\mu \mu_0}
    \label{eq:It0}
.\end{equation}

If $\tau_0 \ll 1$, only single scattered photons can be taken into account. Substituting Eq. \ref{eq:S_inc_2} into Eq. \ref{eq:It0} yields:
\begin{equation}
    I_0 = \frac{\lambda F}{4 \pi} x(\cos{\gamma_0}) \int_{0}^{\tau_0} e^{-\tau\left(\frac{1}{\mu_0} + \frac{1}{\mu}\right)} \frac{d \tau}{\mu \mu_0}
    \label{eq:It0_2}
.\end{equation}
The integral in Eq. \ref{eq:It0_2} can be calculated to give:
\begin{equation}
    \int_{0}^{\tau_0} e^{-\tau\left(\frac{1}{\mu_0} + \frac{1}{\mu}\right)} \frac{d \tau}{\mu \mu_0} = \frac{\mu_0}{\mu_0 + \mu} \left(1 - e^{-\tau_0\left(\frac{1}{\mu} + \frac{1}{\mu_0}\right)}\right)
    \label{eq:int_dt}
.\end{equation}
After expanding the exponent on the right-hand side of Eq. \ref{eq:int_dt} in series, and substituting it to the Eq. \ref{eq:It0_2} we obtain
\begin{equation}
    I_0 = \frac{\lambda}{4\pi} \frac{F \tau_0}{\mu} x(\cos{\gamma_0})\,.
\end{equation}
A useful quantity is the specific luminosity due to scattering: 
\begin{equation}
    \frac{d L}{d \Omega}  \mathrm{\left[\frac{erg}{s\, sr}\right]} = \int I_0 \mu ds
    \label{dldomega}
,\end{equation}
where the integral is calculated over the visible surface of the star. Usually, this integral is computed numerically. 

\begin{figure}
        \includegraphics[width=\columnwidth]{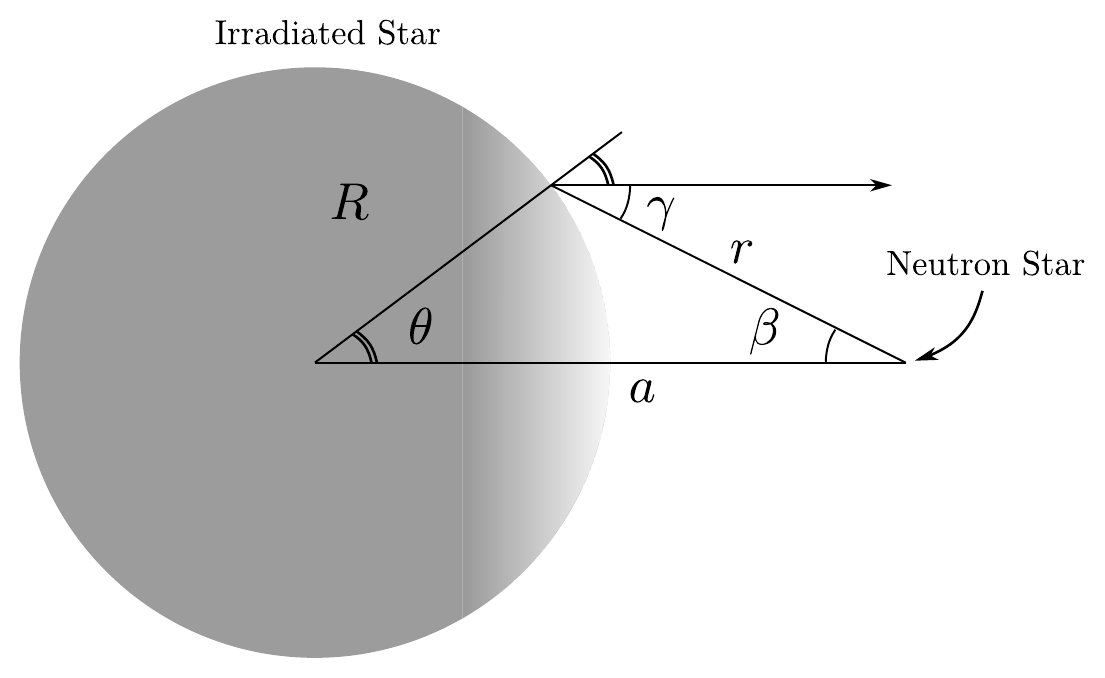}
    \caption{Spherical star illuminated by an isotropic point-like X-ray source. $R$ is the radius of the star, $a$ is the distance between star center and X-ray source, and $r$ is the distance between X-ray source and surface of the star. The observer is located to the right-hand side of the figure. } 
    \label{fig:sperical_star}
\end{figure}

As an example, consider the simple case of a spherical star with radius $R$ illuminated by an isotropic point-like X-ray source located at the distance $a$ from the star center (see Fig. \ref{fig:sperical_star}):
\begin{equation}    
    F = \frac{L_{\rm x}}{4 \pi r^2}    
.\end{equation}
Here the distance $r$ is defined by the cosine theorem:
\begin{equation}
    r^2 = a^2 + R^2 - 2 a R \cos{\theta}
.\end{equation}
The surface element on the star is:
\begin{equation}
    dS = R^2 d\Omega = R^2 \sin{\theta} \, d\theta \, d\phi
.\end{equation}
Then Eq. (\ref{dldomega}) turns into:
\begin{equation}
    \frac{d L}{d \Omega} \mathrm{\left[\frac{erg}{s\, sr}\right]} = \frac{L_x}{4 \pi} \frac{R^2}{a^2} \frac{\lambda \tau_0}{2} \int_{0}^{\theta^{*}} x(\cos{\gamma_0})  \frac{\sin{\theta} d\theta}{\frac{5}{4} - \cos{\theta}}
    \label{eq:Q}
.\end{equation}

For instance, in the case of spherically symmetric source $x(\cos{\gamma_0}) = 1$, the integral in (\ref{eq:Q}) is simplified:
\begin{equation}
     \int_{0}^{\theta^{*}} \frac{\sin{\theta} d\theta}{\frac{5}{4} - \cos{\theta}}\,. 
    \label{eq:int2}
\end{equation}
For $\theta^{*} = 30\degree$ and $R/a = 1/2$, the integral (\ref{eq:int2}) is $=\ln{3} \approx 1.0986$.

In the case of the Rayleigh scattering 
\begin{equation}
   x(\cos{\gamma}) = \frac{3}{4}(1 + \cos^{2}{\gamma})\,.
        \label{eq:x}
\end{equation}
We can use the sine theorem
\begin{equation}
    \frac{R}{\sin{\beta}} = \frac{a}{\sin{(\beta+\theta)}}
\end{equation}
to find the relation between the scattering angle $\gamma$ and $\theta,$
\begin{equation}
    \sin^2{\gamma} = \frac{1}{\left(1 + \frac{R \sin{\theta}}{a - R\cos{\theta}}\right)^2}\,.
\end{equation}
The integral in equation \ref{eq:Q} becomes
\begin{equation}
    \int_{0}^{\theta^{*}} \frac{3}{4} \left(2 - \frac{1}{\left( 1 + \frac{R\sin{\theta}}{a - R\cos{\theta}}\right)^2} \right) \frac{\sin{\theta} d\theta}{\left(\frac{5}{4} - \cos{\theta}\right)}
    \label{eq:int}
.\end{equation}
For $\theta^{*} = 30\degree$ and $R/a = 1/2$, the integral \ref{eq:int} is $\approx 1.24$. The difference between A14 and A17 is approximately 20\%.

X-ray flux scattered by the corona at the distance $d$ to the system at a given orbital phase is 
\begin{equation}
    q' \left[\frac{\mathrm{erg}}{\mathrm{s\,cm^2}}\right] = \frac{1}{d^2} \frac{d L}{d \Omega}
.\end{equation}
The observable X-ray flux from the neutron star at the Main-on phases of 35-day cycle is
\begin{equation}
    q_x = \frac{L_x}{4 \pi d^2}
.\end{equation}
Therefore, the fraction of the X-ray flux scattered by the corona is $q'/q_{\rm x}=4\pi (dL/d\Omega)/L_{\rm x}$.

The scattered flux $q'$ is approximately equal to the flux observable by \textit{eROSITA} in the low state of Her X-1 (the power-law component at $\sim$ 1 keV). The radiation flux from the central source $q_{\rm x}$ is approximately equal to the flux observable during the Main-on; see, e.g., \cite{1997A&A...327..215O, 1998A&A...329L..41D} (the power-law component at $\sim$ 1 keV). Thus, the fraction of scattered radiation in the hot corona can be estimated from observations to be  $q'/q_{\rm x} \approx$ 0.01--0.02. 

Assuming pure scattering ($\kappa=0$, $\lambda=1$, $\tau_0 = \tau_{\sigma}$) and taking $q'/q_{\rm x} = 0.01$, from Eq. (\ref{eq:Q}) we find the scattering optical depth: 
\begin{equation}
    \tau_{\sigma} = \frac{q'}{q_x} \frac{2 a^2}{R^2} \frac{1}{1.24} \approx \frac{1}{15}\,.
\end{equation}

\section{Scattering and absorption of X-ray photons in a semi-infinite atmosphere}

\label{app2}
Theory of radiation transfer with electron scattering was developed in the classical works of
\cite{ambartsumian1960}, \cite{chandra1989} and \cite{sobolev1985}. \cite{1969SvA....13..175Z},  \cite{1972SvA....16..532S} and \cite{1973A&A....24..337S} studied the effect of electron scattering on the spectrum emitted by an isothermal semi-infinite atmosphere. Such a spectrum is referred to as the  ``modified blackbody'' (see, e.g., \cite{1983bhwd.book.....S}, \cite{1986rpa..book.....R}, \cite{2002apa..book.....F} and \cite{2008bhad.book.....K}). The integral radiation flux from photospheres with electron scattering was calculated by \cite{1974SvA....18...60S}. Following the classical papers, the solution of the radiation transfer equation (\ref{eq:transfer}) with the source function \ref{eq:S} reads:
\begin{equation}
    I(\tau = 0, \mu, \mu_0) = F \rho(\mu, \mu_0) \mu_0
,\end{equation}
where $\rho(\mu, \mu_0)$ is the atmosphere brightness coefficient:
\begin{equation}
    \rho(\mu, \mu_0) = \frac{\lambda}{4} \left( x_0 \frac{\phi_0(\mu) \phi_0(\mu_0)}{\mu + \mu_0} + x_2 \frac{\phi_2(\mu) \phi_2(\mu_0)}{\mu+\mu_0} \right).
\end{equation}
Following \cite{busbridge1960} $\phi_i(\mu)$ is defined as:
\begin{multline}
    \phi_0(\mu) = H(\mu) q_0(\mu),\\
    \phi_2(\mu) = H(\mu) q_2(\mu),\\
\end{multline}
where $q_0(\mu)$ and $q_2(\mu)$ are the following polynomials \citep{1968SvA....12..420S}:
\begin{multline}
    q_0(\mu) = 1 + \frac{N_2 M_0 - M_2 N_0}{\Delta} \mu + \frac{M_1 N_0 - N_1 M_0}{\Delta} \mu^2\\
    q_2(\mu) = -\frac{1}{2} q_0(\mu) - \frac{3(1 - \lambda)}{2\Delta} (M_2\mu - M_1\mu^2).\\
\end{multline}
Here $M_0, M_1, M_2, N_0, N_1, \Delta$ coefficients are defined as:
\begin{multline}
    M_0 = - \frac{\lambda}{2} (1-\lambda) \int_0^1 H(\mu) \left[ 3 \frac{x_2}{2} P_2(\mu) \right] \mu d\mu,\\
    M_1 = 1 - \frac{\lambda}{2} \int_0^1 H(\mu) \left[1 - \frac{x_2}{2} P_2(\mu) \right] d\mu,\\
    M_2 = - \frac{\lambda}{2} \int_0^1 H(\mu) \left[1 - \frac{x_2}{2} P_2(\mu) \right] \mu d\mu,\\
    N_0 = - \frac{\lambda}{4} x_2 3(1-\lambda) \int_0^1 H(\mu) P_2(\mu) d\mu,\\
    N_1 = \frac{\lambda}{4} x_2 3(1-\lambda) \int_0^1 H(\mu) P_2(\mu) \mu d\mu,\\
    N_2 = 1 - \frac{\lambda}{2} \int_0^1 H(\mu) \left[1 - \frac{x_2}{2} P_2(\mu) \right] d\mu,\\
    \Delta = M_1 N_2 - M_2 N_1,\\
\end{multline}
where $P_2(\mu)$ is the 2d-order Legendre polynomial. For the Rayleigh scattering diagram \ref{eq:x},  $N_2 = M_1$.

The function $H(\mu)$ is defined as  \citep{1950ratr.book.....C}:
\begin{equation}
    H(\mu) = 1 + \mu H(\mu) \int_0^1 \frac{\psi(\mu)H(\mu')}{\mu + \mu'} d\mu'\, ,
\end{equation}
where $\psi(\mu)$ is the characteristic function:
\begin{equation}
\label{f:psi}
    \psi(\mu) = \frac{\lambda}{2} \left[ 1 + \frac{x_2}{2}(3(1-\lambda)\mu^2 - 1)P_2(\mu) \right] \,.
\end{equation}
Here $x_2 = 0$ for isotropic scattering, $x_2 = 1/2$ for the Rayleigh scattering \citep{1968SvA....12..420S}. The characteristic function \ref{f:psi} satisfies the condition:
\begin{equation}
    \int_0^1 \psi(\mu)d\mu \leq \frac{1}{2}
.\end{equation}

X-ray flux scattered by the photosphere at the distance $d$ to the system at a given orbital phase is
\begin{equation}
    q_{\rm phot} = \frac{L_{\rm{x}}}{4\pi d^2} \int \frac{\rho(\mu, \mu_0) \mu \mu_0}{r^2} ds,
\end{equation}
where the integral is taken over the visible part of the irradiated surface of the donor star. 

\begin{figure}
        \includegraphics[width=\columnwidth]{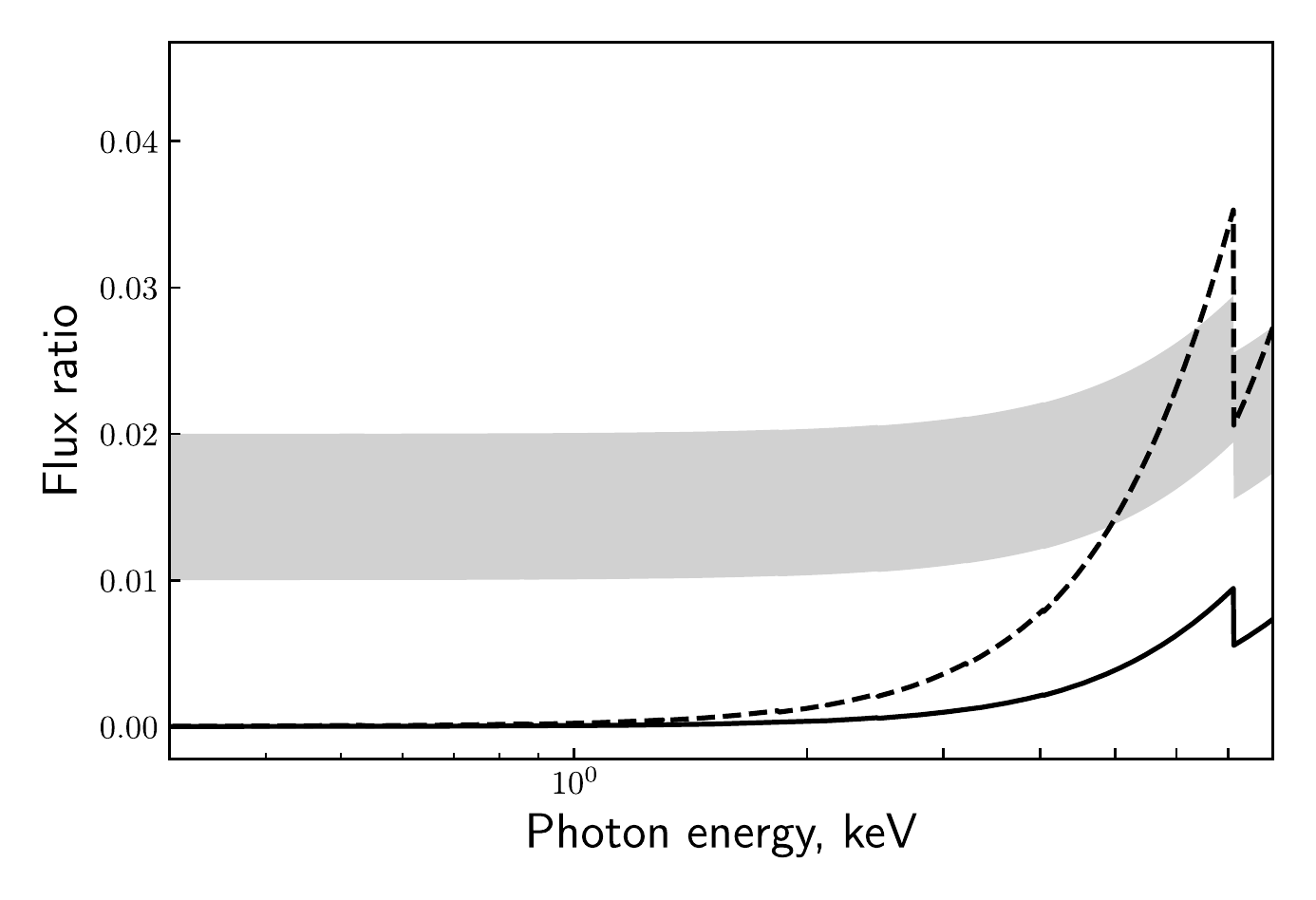}
    \caption{Ratio of the continuum X-ray fluxes at low state to Main-on state as a function of photon energy. A fluorescent iron line is not included. The dashed and solid lines show the calculated fraction of scattered X-ray radiation from the photosphere $q_{\rm phot}/q_{\rm x}$ without and with taking into account the shadow from the disk, respectively. The gray area shows the observable $(q' + q_{\rm phot})/q_{\rm x}$ fraction with an addition of X-rays scattered in the corona ($\sim$ 0.01--0.02).     }
    \label{fig:qq}
\end{figure}

In Fig. \ref{fig:qq}, we plot the scattered flux ratio $q_{\rm phot}/q_{\rm x}$ from the cold photosphere of HZ Her as a function of the photon energy (the dashed and solid curves for calculation without and with taking into account the disk shadow, respectively). Besides, the gray strip shows the total ratio of scattered flux from the hot corona $q'$ and the reflected flux from the cold photosphere $q_\mathrm{phot}$ to the flux from the central X-ray source $q_{\rm x}$. The low and upper contributions from the scattering corona $q^{\prime}_{\rm min}/q_{\rm x}=0.01$ and $q^{\prime}_{\rm max}/q_{\rm x}=0.02$ are used (see Appendix A).

We note that the theory described above was applied for the first time by \citet{2011AstL...37..311M} to calculate the vertical structure of accretion disks.

\end{appendix}

\end{document}